\documentclass[12pt]{article}
\usepackage{graphicx}
\def\e{{\rm E}}
\def\apj{{\it Astrophys. J. \ }}

\def\mnras{{\it Mon. Not. R. Astron. Soc. \ }}

\def\nat{{\it Nature \,}}
\usepackage{scicite}

\usepackage{times}
\usepackage{scicite}
\usepackage{psfig}

\newenvironment{sciabstract}{%
\begin{quote} \bf}
{\end{quote}}

\newcounter{lastnote}

\newcommand{\mathbold}[1]{\mbox{\boldmath $\bf#1$}}

\title{
Discovery of a Jupiter/Saturn Analog with Gravitational Microlensing
}

\author{
B.S.~Gaudi$^{1,\ast}$,
D.P.~Bennett$^{2}$, 
A.~Udalski$^{3}$,
A.~Gould$^{1}$,\\
G.W.~Christie$^{4}$,
D.~Maoz$^{5}$,
S.~Dong$^{1}$, 
J.~McCormick$^{6}$,\\
M.K.~Szyma{\' n}ski$^{3}$, 
P.J.~Tristram$^{7}$, 
S.~Nikolaev$^{8}$,\\
B.~Paczy{\' n}ski$^{9,\dagger}$,
M.~Kubiak$^{3}$, 
G.~Pietrzy{\' n}ski$^{3,10}$, 
I.~Soszy{\' n}ski$^{3}$,\\
O.~Szewczyk$^{3}$,
K.~Ulaczyk$^{3}$, 
{\L}.~Wyrzykowski$^{3,11}$\\ 
(The OGLE Collaboration)\\
D.L.~DePoy$^{1}$, 
C.~Han$^{12}$, 
S.~Kaspi$^{5}$, 
C.-U.~Lee$^{13}$, 
F.~Mallia$^{14}$,\\
T.~Natusch$^{4}$,
R.W.~Pogge$^{1}$, 
B.-G.~Park$^{13}$, 
(The $\mu$FUN Collaboration)\\
F.~Abe$^{15}$, 
I.A.~Bond$^{16}$, 
C.S.~Botzler$^{17}$, 
A.~Fukui$^{15}$, 
J.B.~Hearnshaw$^{18}$,\\
Y.~Itow$^{15}$,  
K.~Kamiya$^{15}$, 
A.V.~Korpela$^{19}$,
P.M.~Kilmartin$^{7}$, 
W.~Lin$^{16}$,\\
K.~Masuda$^{15}$,  
Y.~Matsubara$^{15}$, 
M.~Motomura$^{15}$, 
Y.~Muraki$^{20}$, 
S.~Nakamura$^{15}$,\\
T.~Okumura$^{15}$, 
K.~Ohnishi$^{21}$, 
N.J.~Rattenbury$^{22}$, 
T.~Sako$^{15}$, 
To.~Saito$^{23}$,\\
S.~Sato$^{24}$, 
L.~Skuljan$^{16}$, 
D.J.~Sullivan$^{19}$, 
T.~Sumi$^{15}$, 
W.L.~Sweatman$^{16}$,\\
P.C.M.~Yock$^{17}$, (The MOA Collaboration)\\
M.D.~Albrow$^{18}$, 
A.~Allan$^{25}$, 
J.-P.~Beaulieu$^{26}$, 
M.J.~Burgdorf$^{27}$, 
K.H.~Cook$^{8}$, \\
C.~Coutures$^{26}$,
M.~Dominik$^{28,\ddag}$, 
S.~Dieters$^{29}$, 
P.~Fouqu\'e$^{30}$, 
J.~Greenhill$^{29}$,\\
K.~Horne$^{28}$,
I.~Steele$^{27}$, 
Y.~Tsapras$^{27}$,\\
(From the PLANET and RoboNet Collaborations)\\
B.~Chaboyer$^{31}$, 
A.~Crocker$^{32}$, 
S.~Frank$^{1}$, 
B.~Macintosh$^{8}$\\
}

\begin{document}


\maketitle 

\noindent
\normalsize{$^{1}$Department of Astronomy, Ohio State University, 140 West 18th Avenue, Columbus, OH 43210, USA}\\
\normalsize{$^\ast$To whom correspondence should be addressed; E-mail: gaudi@astronomy.ohio-state.edu}\\
\normalsize{$^{2}$Department of Physics, 225 Nieuwland Science Hall, Notre Dame University, Notre Dame, IN 46556, USA}\\
\normalsize{$^{3}$Warsaw University Observatory, Al.~Ujazdowskie~4, 00-478~Warszawa, Poland}\\
\normalsize{$^{4}$Auckland Observatory, P.O. Box 24-180, Auckland, New Zealand}\\
\normalsize{$^{5}$School of Physics and Astronomy, Raymond and Beverley Sackler Faculty of Exact Sciences, Tel-Aviv University, Tel Aviv 69978, Israel}\\
\normalsize{$^{6}$Farm Cove Observatory, 2/24 Rapallo Place, Pakuranga, Auckland 1706, New Zealand}\\
\normalsize{$^{7}$Mt. John Observatory, P.O. Box 56, Lake Tekapo 8770, New Zealand}\\
\normalsize{$^{8}$IGPP, Lawrence Livermore National Laboratory, 7000 East Ave., Livermore, CA 94550, USA}\\
\normalsize{$^{9}$Princeton University Observatory, Princeton, NJ 08544, USA}\\
\normalsize{$^\dagger$Deceased}\\
\normalsize{$^{10}$Universidad de Concepci{\'o}n, Departamento de Fisica, Casilla 160--C, Concepci{\'o}n, Chile}\\
\normalsize{$^{11}$Institute of Astronomy, University of Cambridge, Madingley Road, Cambridge CB3 0HA, UK}\\
\normalsize{$^{12}$Program of Brain Korea, Department of Physics, Chungbuk National University, 410 Seongbong-Rho, Hungduk-Gu, Chongju 371-763, Korea}\\
\normalsize{$^{13}$Korea Astronomy and Space Science Institute, 61-1 Hwaam-Dong, Yuseong-Gu, Daejeon 305-348, Korea}\\
\normalsize{$^{14}$Campo Catino Astronomical Observatory, P.O. Box Guarcino, Frosinone 03016, Italy}\\
\normalsize{$^{15}$Solar-Terrestrial Environment Laboratory, Nagoya University, Nagoya, 464-8601, Japan}\\
\normalsize{$^{16}$Institute for Information and Mathematical Sciences, Massey University, Private Bag 102-904, Auckland 1330, New Zealand}\\
\normalsize{$^{17}$Department of Physics, University of Auckland, Private Bag 92-019, Auckland 1001, New Zealand}\\
\normalsize{$^{18}$University of Canterbury, Department of Physics and Astronomy, Private Bag 4800, Christchurch 8020, New Zealand}\\
\normalsize{$^{19}$School of Chemical and Physical Sciences, Victoria University, Wellington, New Zealand}\\
\normalsize{$^{20}$Department of Physics, Konan University, Nishiokamoto 8-9-1, Kobe 658-8501, Japan}\\
\normalsize{$^{21}$Nagano National College of Technology, Nagano 381-8550, Japan}\\
\normalsize{$^{22}$Jodrell Bank Centre for Astrophysics, The University of Manchester, Manchester, M13 9PL, UK}\\
\normalsize{$^{23}$Tokyo Metropolitan College of Aeronautics, Tokyo 116-8523, Japan}\\
\normalsize{$^{24}$Department of Physics and Astrophysics, Faculty of Science, Nagoya University, Nagoya 464-8602, Japan}\\
\normalsize{$^{25}$School of Physics, University of Exeter, Stocker Road, Exeter, EX4 4QL, UK}\\
\normalsize{$^{26}$Institut d'Astrophysique de Paris, CNRS, Université Pierre et Marie Curie UMR7095, 98bis Boulevard Arago, 75014 Paris, France}\\
\normalsize{$^{27}$Astrophysics Research Institute, Liverpool John Moores University, Twelve Quays House, Egerton Wharf, Birkenhead CH41 1LD, UK}\\
\normalsize{$^{28}$SUPA, University of St Andrews, School of Physics \& Astronomy,North Haugh, St Andrews, KY16 9SS, UK}\\
\normalsize{$^\ddag$Royal Society University Research Fellow}\\
\normalsize{$^{29}$University of Tasmania, School of Mathematics and Physics, Private Bag 37, Hobart, TAS 7001, Australia}\\
\normalsize{$^{30}$Observatoire Midi-Pyr\'en\'ees, Laboratoire d'Astrophysique, UMR 5572, Universit\'{e} Paul Sabatier--Toulouse 3, 14 avenue Edouard Belin, 31400 Toulouse, France}\\
\normalsize{$^{31}$Department of Physics and Astronomy, Dartmouth College, 6127 Wilder Laboratory, Hanover, NH 03755, USA}\\
\normalsize{$^{32}$University of Oxford, Denys Wilkinson Building, Keble Road, Oxford, OX1 3RH, UK}\\



\begin{sciabstract}
Searches for extrasolar planets have uncovered an astonishing
diversity of planetary systems, yet the frequency of solar system
analogs remains unknown.  The gravitational microlensing planet search
method is potentially sensitive to multiple-planet systems containing
analogs of all the solar system planets except Mercury.  We report the
detection of a multiple-planet system with microlensing.  We identify
two planets with masses of $\mathbold{\sim 0.71}$ and $\mathbold{\sim 0.27 }$ 
times the mass of Jupiter and orbital separations of $\mathbold{\sim 2.3}$ and 
$\mathbold{\sim 4.6}$ astronomical units orbiting a primary star
of mass $\mathbold{\sim 0.50}$ solar masses at a distance of
$\mathbold{\sim 1.5}$ kiloparsecs.   This system resembles a scaled
version of our solar system in that the mass ratio, separation ratio,
and equilibrium temperatures of the planets are similar to those
of Jupiter and Saturn.  These planets could not have been detected
with other techniques; their discovery from only six confirmed
microlensing planet detections suggests that solar system analogs
may be common.
\end{sciabstract}


Nearly 250 extrasolar planets \cite{encyclopedia} have been discovered
by measuring a variety of effects: reflex motion of the host star
using pulsar timing or precision Doppler measurements
\cite{pulsar,51peg,70vir}; periodic dimming of the parent star as the
planet transits in front \cite{udalski,konacki}; and planet-induced
perturbations to microlensing light curves in which the host star acts
as the primary gravitational lens
\cite{mp91,ob03235,ob05071,ob05390,ob05169}.  These detections have
uncovered an enormous range of planetary properties, indicating that
planetary systems very unlike our own are common throughout the Galaxy
\cite{fischer}.

To date, $\sim 25$ multiple-planet systems have been detected
\cite{catalog}, all but one \cite{pulsar} using the Doppler method.
Because Doppler surveys must monitor the host star's reflex motion over
the planet's orbital period, they are limited by the finite duration
as well as the sensitivity of the measurements.  Hence, they are only
just now becoming sensitive to Jupiter analogs and are not yet
sensitive to Saturn analogs (nor, ipso facto, Jupiter/Saturn systems).
Thus, all multiple-planet systems discovered so far are very
dissimilar from our own, and the frequency of solar system analogs
remains unknown.

Because microlensing relies on the direct perturbation of light from
distant stars by the gravitational field of the planet, it is
`instantaneously' able to detect planets without requiring
observations over a full orbit.  For a primary star of mass $M$,
microlensing sensitivity peaks for planets in the range $\sim
[1-5](M/0.3 M_\odot)^{1/2}$ astronomical units (AU)
\cite{gouldloeb92}.  For solar-mass stars, this is exactly the range
of the solar system gas giants so microlensing provides a method to
probe solar system analogs \cite{gouldloeb92,bennettrhie02}.

As pointed out by Griest \& Safizadeh \cite{griestsafi}, the very rare
class of high-magnification ($>100$) microlensing events provides an
extremely sensitive method of detecting planets.  Near the peak of
high-magnification events, the two images created by the primary star
are highly magnified and distorted, and form a complete or nearly
complete Einstein ring.  A planetary companion to the primary star
lying reasonably near the Einstein ring will distort the symmetry of
the ring.  As the host passes very close to the source line-of-sight,
the images sweep around the Einstein ring, thus probing this
distortion.  Although the total number of high-magnification events is
small, the instantaneous chance of detection in each is much higher
than for the more common low-magnification events.  Equally important,
the interval of high-sensitivity (i.e., high-magnification) is
predictable from the evolution of the light curve
\cite{griestsafi,rhie00,rattenbury02,abe04}.  This permits
concentration of scarce observing resources on these events.
Furthermore, the high-magnification makes it possible to acquire high
signal-to-noise ratio photometry of the peak of the events using
relatively small-aperture (and so plentiful) telescopes.  As a result,
four \cite{ob05071,ob05169} of the six planets \cite{ob03235,ob05390}
discovered to date in microlensing events were in high-magnification
events.

Almost immediately after Griest and Safizadeh \cite{griestsafi} pointed out the sensitivity
of high-magnification events, Gaudi et al.\ \cite{gaudi98} derived an important
corollary.  Because planets in the neighborhood of the Einstein ring are
revealed with near unit probability in high-magnification events,
multiple-planet systems lying in this region will be revealed with
almost the same probability.

The Optical Gravitational Lens Experiment (OGLE) \cite{ews} and
Microlensing Observations in Astrophysics (MOA) \cite{abe04}
collaborations together alert $\sim 700$ ongoing microlensing events
per year.  Two collaborations, a joint venture of the Probing Lensing
Anomalies NETwork (PLANET) \cite{albrow98} and RoboNet
\cite{burgdorf07} collaborations, and the Microlensing Follow-Up
Network ($\mu$FUN) \cite{yoo04}, then monitor a subset of these alerts
to search for planets.  $\mu$FUN focuses almost entirely on
high-magnification events, including two events originally alerted by
OGLE that proved to have a Jupiter-mass \cite{ob05071} and a
Neptune-mass \cite{ob05169} planet, respectively.  Here, we report on
the detection of a multi-planet system using this approach.

On 28 March 2006 (HJD$\sim3822$), the OGLE Early Warning System (EWS)
\cite{ews} announced OGLE-2006-BLG-109 
as a non-standard microlensing event possibly indicative of a planet.
This immediately triggered followup observations by $\mu$FUN
and RoboNet, which gained intensity as the event approached
high-magnification.  On 5 April, the event underwent a deviation from
the single-lens form indicative of a binary lens. Within 12 hours of
this deviation, a preliminary model indicated a jovian-class planet,
which was predicted to generate an additional peak on 8 April.  The 8
April peak occurred as predicted, but in the meantime, there was an
additional peak on 5/6 April, which turned out to be due to a second
Jovian-class planet.

Figure~1 shows the data from 11 observatories, including 7 from
$\mu$FUN [the Auckland 0.35m and Farm Cove 0.25m in New Zealand (clear
filter), the Wise 1m in Israel (clear), the CTIO/SMARTS 1.3m in
Chile ($I$-band and $H$-band), the Areo8 0.3m in New Mexico operated
by the Campo Catino Astronomical Observatory (clear), and the MDM
2.4m ($I$-band) and Mt. Lemmon 1.0m ($I$-band) in Arizona], the OGLE
Warsaw 1.3m ($I$-band) in Chile, the MOA Mt.\ John 0.6m ($I$-band) in
New Zealand, the PLANET/Canopus 1m ($I$-band) in Tasmania, and the
RoboNet/Liverpool 2m ($R$-band) in the Canary Islands.  There are a
total of 2642 data points. In addition, there are 29 V-band data
points from OGLE and CTIO/SMARTS that we use to determine the source
color.

The qualitative character of the event can be read directly from the
light curve, primarily from the five distinctive features shown in Figure
2.  Consider the first three features: the low-amplitude anomaly (OGLE,
HJD$\sim 3823$) that triggered the OGLE EWS alert, the gentle
``shoulder'' during the first rise (MOA, HJD$\sim 3830$), and the
first peak (Auckland, HJD$\sim 3831$).  Together, these can only be
produced by, respectively, passage close to or over a weak cusp,
entrance into a weak caustic, and exit from a strong caustic.  (The
magnification diverges when a point source crosses a closed concave
caustic curve, where additional images are created on entry or
destroyed on exit of the enclosed area.  Caustics are strong or weak
depending on the brightness of these images.  The concave curves meet
at cusps that produce sharp spikes of magnification when crossed by
the source.) Such a sequence requires a topology similar to the one
shown in the inset to Figure 1.  The specific strengths of each
feature require the specific caustic topology shown.  In particular,
the narrow mouth of the caustic toward the bottom generates a very
strong caustic.  This was essentially the argument used to predict the
fifth feature (OGLE/MDM/Lemmon/Auckland/FarmCove/Tasmania, HJD$\sim 3834$),
corresponding to a moderately strong cusp passage (Fig.~1).  The size
and strength of this caustic imply a jovian-class planet lying very
close to the Einstein ring, although detailed modeling is required to
derive the precise planet/star mass ratio.  The fourth feature
(Wise/OGLE, HJD$\sim 3831.5$) cannot be explained by considering the
caustic generated by this jovian-class planet alone.  This feature
occurs near the time when the source approaches closest to the
center-of-mass of the planet/star system; this is exactly the time at
which the central-caustic bumps due to additional planets are expected
to occur \cite{gaudi98}.  The inset in Figure 1 highlights the
additional caustic feature due to a second planet that is required to
explain this bump.  This caustic feature is smaller than the main
caustic, which implies that the planet, also of jovian class, lies
farther from the Einstein ring, so it is subject to the standard
\cite{griestsafi} $b\leftrightarrow b^{-1}$ degeneracy, where $b$ is
the planet-star projected separation in units of the Einstein radius.
A detailed analysis shows the mass is three times as great as
that of the
first planet and that the $b<1$ solution is favored by
$\Delta\chi^2=11.4$.  We label these planets OGLE-2006-BLG-109Lc and
OGLE-2006-BLG-109Lb, respectively.  Although the caustics of the
individual planets do interact to form a single caustic curve, their
effects are nevertheless mostly independent
\cite{wambsganss97,rattenbury02,han05}, so the parts of the caustic
associated with the individual planets can be identified, as shown in
Figure 1.  Modeling the light curve in detail with a three-body lens
yields, $m_b/M = 1.35\times 10^{-3}$, $m_c/m_b = 0.36$ for the mass
ratio of the planets and their host, very similar to $m_j/M_\odot =
0.96\times 10^{-3}$ and $m_s/m_j= 0.30$ for Jupiter, Saturn, and the
Sun.  The ratio of projected separations $r_{\perp,b}/r_{\perp,c}=
0.60$ is also very similar to the Jupiter/Saturn value of $a_j/a_s =
0.55$.

Two subgroups of authors conducted independent searches for
alternative solutions.  Both found that no single-planet solution is
consistent with the light-curve topology.  We also successively
eliminated each of the five features to see whether the remaining four
features could be fit by a single planet.  We found that only
the elimination of the fourth feature produced successful single-planet
models.  By contrast, similar procedures in other events
\cite{ob05169,dong07} led to many independent solutions.
OGLE-2006-BLG-109 differs from these in that it has five well-covered
features.

Several higher-order effects are apparent in this event that
permit us to extract much more detailed information about the
system from the light curve.  We only briefly sketch these here.  For
over 95\% of microlensing events observed from the ground, the lens
parameters are determined only relative to the angular Einstein radius
$\theta_\e$, whose absolute scale remains unknown.  Here
$\theta_\e=\sqrt{4GM/c^2D}$, where $M$ is the mass of the lens,
$1/D \equiv 1/D_l-1/D_s$, and $D_l$ and $D_s$ are the distances to the lens
and the source, respectively.  However, in this event the effect of the
finite size of the source star during caustic exit allows us to
measure the source radius relative to the Einstein radius,
$\rho=\theta_*/\theta_\e$ \cite{gould94}.  From the source color and
flux we can determine its angular size $\theta_*$, and thus
$\theta_\e$ \cite{yoo04}.

The acceleration of Earth in its orbit about the Sun induces subtle
distortions on the light curve called microlens parallax, which yields
the physical size of the Einstein radius projected onto the observer
plane, $\tilde r_\e \equiv \theta_\e D$ \cite{gould92}.  This is
usually measured only in the roughly 3\% of events that are extremely
long, but this event happens to be long and so displays clear
distortions arising from parallax.

Combining these two measures of the Einstein radius allows us to
triangulate the event and so determine the host star distance, $D_l =
1/(\theta_\e/\tilde r_\e + 1/D_s)$, and mass, $M=(c^2/4 G)\tilde
r_\e\theta_\e$.  We assume $D_s = 8\,$kpc, although our results are
insensitive to this assumption. From a preliminary analysis we
infer $D_l \simeq 1.5\, $kpc and $M \simeq 0.5\, M_\odot$.  Based on
high-resolution Keck AO $H$-band images, we detect light from the lens
and infer its magnitude to be $H=17.17 \pm 0.25$, consistent with the
mass estimate from the light curve.  We subsequently incorporate the
lens flux constraint in the light curve analysis, which allows us to
derive more precise estimates of $D_l = 1.49 \pm 0.13~{\rm kpc}$ and
$M=0.50 \pm 0.05~M_\odot$.  The planet masses are $m_b=0.71\pm0.08$
and $m_c= 0.27 \pm 0.03$ times the mass of Jupiter.

Finally, we also detect the orbital motion of the outer planet; this
motion both rotates and changes the shape of the larger caustic shown
in the top inset to Figure 1.  We are able to constrain the two
components of the projected velocity of the planet relative to the
primary star.  Together with the estimate of the stellar mass, they
completely determine the outer planet's orbit (including inclination)
under the assumption that it is circular, up to a two-fold degeneracy.
The solution presented here is marginally favored by the data at $\Delta\chi^2=4.8$
via the effect of the planet's acceleration on the light curve.
Thus we can estimate the full (three-dimensional) separation of planet
c (again assuming a circular orbit), and also of planet b (assuming a
coplanar and circular orbit).  We find $a_b=2.3\pm 0.2~{\rm AU}$ and
$a_c=4.6\pm 0.5~{\rm AU}$.  A more refined estimate of these
parameters and their uncertainties will require a detailed analysis
including the combined effects of finite sources, parallax, and
orbital motion of the planets. The results of this analysis will be
presented elsewhere (Bennett et~al., in preparation).

The OGLE-2006-BLG-109L planetary system bears a remarkable similarity
to our own solar system.  Although the primary mass is only half solar,
the mass ratio of the two planets ($0.37$) and separation ratio
($0.50$) are similar to those of Jupiter and Saturn.  We infer their
equilibrium temperatures to be $T_{\rm eq}\sim 82 \pm 12~{\rm K}$
and $T_{\rm eq}\sim 59 \pm 7~{\rm K}$, $\sim 30\%$ smaller than
those of Jupiter and Saturn.

Before the detection of extrasolar planets, planet formation
theories generally predicted that other systems should resemble our
solar system.  In the core-accretion paradigm, the most massive giant
planet forms at the `snow line,' the point in the protoplanetary disk
exterior to which ices are stable.  Immediately beyond the snow line,
the surface density of solids is highest and the dynamical time is the
shortest, and therefore the timescale for planet formation is the
shortest. Beyond the snow line, the formation timescale increases with
distance from the host star.  Thus in this `classical' picture of
planet formation, one would expect planet mass to decrease with
increasing distance beyond the snow line, as is observed in our solar
system \cite{lissauer}. The discovery of a population of massive
planets well interior to the snow line demonstrated that this picture
of planet formation is incomplete, and considerable inward migration of
planets must occur \cite{lin96}.  Nevertheless, this classical picture
may still be applicable to our solar system and some fraction of other
systems as well.  The OGLE-2006-BLG-109L planetary system represents
just such a `scaled version' of our own solar system, with a
less-massive host.  This system preserves the mass-distance
correlation in our solar system, and the scaling with primary mass is
consistent with the core-accretion paradigm in which giant planets
that form around lower-mass stars are expected to be less massive but
form in regions of the protoplanetary disk with similar equilibrium
temperatures and are therefore closer to their parent star \cite{idalin}.

The majority of the $\sim 25$ known multi-planet systems are quite
dissimilar to the OGLE-2006-BLG-109L system and to our own solar
system.  Many of these systems have the very close-in massive planets
indicative of large-scale planetary migration, or they have a `normal
hierarchy', in which the masses of the giant planets increase with
distance from the parent star.  There are two multi-planet systems
with properties roughly similar to those of OGLE-2006-BLG-109L.  The
47 UMa and 14 Her systems each contain a giant planet at a semimajor
axis of $\sim 3~{\rm AU}$ and a second, less massive giant planet at
a separation of $\sim 7~{\rm AU}$ \cite{witten}.  However, because of their
higher-mass primaries, the equilibrium temperatures of these planets are
considerably higher than those of OGLE-2006-BLG-109L or Jupiter and
Saturn, so these systems cannot be considered close analogs of our
solar system.

OGLE-2006-BLG-109Lb and OGLE-2006-BLG-109Lc are the fifth and sixth planets to be detected
by microlensing.  Although, given the detection of planet c, the {\it 
a priori} probability of detecting planet b in this event was high,
it was not unity.  Furthermore, only two other jovian-mass planets
have been detected by microlensing \cite{ob03235,ob05071}, and
neither event had substantial sensitivity to multiple planets.  These
facts may indicate that the stars being probed by microlensing that
host jovian-mass companions are also likely to host additional giant
planets.  If the OGLE-2006-BLG-109L planetary system is typical, these
systems may have properties similar to our solar system.  Regardless,
the detection of the OGLE-2006-BLG-109L planetary system demonstrates
that microlensing surveys will be able to constrain the frequency of
solar system analogs throughout the Galaxy.

\bibliography{scibib}

\bibliographystyle{Science}

\end{document}